\documentclass{emulateapj}

\usepackage{psfig}

\usepackage{verbatim}

\newcommand{\etal}{{et~al.}}
\newcommand{\Lsun}{\>{\rm L_{\odot}}}
\newcommand{\spitzer}{\textit{Spitzer Space Telescope}}
\newcommand\bootes{Bo\"{o}tes}
\newcommand\firstobject{Source 1}
\newcommand\firstobjectlong{MIPS J142824.0+352619}
\newcommand\secondobject{Source 2}
\newcommand\secondobjectlong{SST24 J142827.19+354127.71}
\newcommand\lfir{{\rm L}_{8-1000\micron}}
\newcommand\mips{\textit{MIPS}}
\newcommand\irs{\textit{IRS}}
\newcommand\irac{\textit{IRAC}}
\newcommand\lsix{{\rm L}_{{\rm PAH}6.2\micron}}
\bibpunct[;]{(}{)}{;}{a}{}{,}

\shorttitle{\irs \ spectra of two ULIRGs at $\lowercase{z}=1.3$}
\shortauthors{Desai \etal}

\begin{document}

\title{\irs \ spectra of two ultraluminous infrared galaxies at $\lowercase{z} = 1.3$}

\author{V.~Desai\altaffilmark{1}, L.~Armus\altaffilmark{2}, B.~T.~Soifer\altaffilmark{1,2}, D.~W.~Weedman\altaffilmark{3}, S.~Higdon\altaffilmark{3}, C.~Bian\altaffilmark{1}, C.~Borys\altaffilmark{1}, H.~W.~W.~Spoon\altaffilmark{3}, V.~Charmandaris\altaffilmark{4,3,5}, K.~Brand\altaffilmark{6}, M.~J.~I.~Brown\altaffilmark{7}, A.~Dey\altaffilmark{6}, J.~Higdon\altaffilmark{3}, J.~Houck\altaffilmark{3}, B.~T.~Jannuzi\altaffilmark{6}, E.~Le Floc'h\altaffilmark{8,5}, M.~L.~N. Ashby\altaffilmark{9}, H.~A.~Smith\altaffilmark{9}}
\altaffiltext{1}{Division of Physics, Mathematics and Astronomy, 320-47, California Institute of Technology, Pasadena CA 91125}
\altaffiltext{2}{SIRTF Science Center, 314-6, California Institute of Technology, Pasadena CA 91125}
\altaffiltext{3}{Astronomy Department, Cornell University, Ithaca NY 14853}
\altaffiltext{4}{Department of Physics, University of Crete, GR-71003, Heraklion, Greece}
\altaffiltext{5}{Chercheur Associ\'e, Observatoire de Paris, F-75014, Paris, France}
\altaffiltext{6}{National Optical Astronomy Observatory, Tucson AZ 85726-6732}
\altaffiltext{7}{Princeton University Observatory, Peyton Hall, Princeton NJ 08544}
\altaffiltext{8}{Steward Observatory, University of Arizona, Tucson AZ 85721}
\altaffiltext{9}{Harvard-Smithsonian Center for Astrophysics, 60 Garden Street, Cambridge MA  02138}

\begin{abstract}
We present low-resolution ($64 < R < 124$) mid-infrared (8--38
$\micron$) spectra of two $z \approx 1.3$ ultraluminous infrared
galaxies with $\lfir \approx 10^{13} \Lsun$: \firstobjectlong \ and
\secondobjectlong.  The spectra were taken with the \textit{Infrared
Spectrograph} (\irs) \ on board the \spitzer.  Both objects were
discovered in a \textit{Spitzer}/\mips \ survey of the \bootes \ field
of the NOAO Deep Wide-Field Survey (NDWFS).

\firstobjectlong \ is a bright 160 $\micron$ source with a large
infrared-to-optical flux density ratio.  Previous authors provide
evidence for a foreground lens and estimate an amplification of
$\la$10, although this factor is presently poorly constrained.  The
6.2, 7.7, 11.3, and 12.8 $\micron$ PAH emission bands in its \irs \
spectrum indicate a redshift of $z \approx 1.3$.  The large equivalent
width of the 6.2 $\micron$ PAH feature indicates that at least 50\% of
the mid-infrared energy is generated in a starburst, an interpretation
that is supported by a large [\ion{Ne}{2}]/[\ion{Ne}{3}] ratio and a
low upper limit on the X-ray luminosity.

\secondobjectlong \ has the brightest 24 $\micron$ flux (10.55 mJy)
among optically faint ($R > 20$) galaxies in the NDWFS.  Its
mid-infrared spectrum lacks emission features, but the broad 9.7
$\micron$ silicate absorption band places this source at $z \approx
1.3$.  Optical spectroscopy confirms a redshift of $z = 1.293 \pm
0.001$.  Given this redshift, \secondobjectlong \ has among the
largest rest-frame 5 $\micron$ luminosities known.  The similarity of
its SED to those of known AGN-dominated ULIRGs and its lack of either
PAH features or large amounts of cool dust indicate that the powerful
mid-infrared emission is dominated by an active nucleus rather than a
starburst.

Our results illustrate the power of the \irs \ in identifying massive
galaxies in the well-known ``redshift desert'' between $1 < z < 2$ and
in discerning their power sources.  Because they are bright,
\firstobjectlong \ (pending future observations to constrain its
lensing amplification) and \secondobjectlong \ are useful $z > 1$
templates of a high luminosity starburst and AGN, respectively.

\end{abstract}



\keywords{galaxies: formation --- galaxies: evolution --- galaxies: starburst --- galaxies: active}

\section{Introduction}
\label{sec:intro}

Observations at long wavelengths, particularly those made by the
\textit{Infrared Astronomical Satellite} (\textit{IRAS}) and the
\textit{Infrared Space Observatory} (\textit{ISO}), have revealed a
low-redshift population of galaxies which emit most of their
luminosity in the far infrared.
Ultraluminous infrared galaxies
\citep[ULIRGs;][]{Soifer84,Sanders88,SandersMirabel96}, which have
far-infrared luminosities greater than $\lfir = 10^{12}~{\rm
L}_{\odot}$, represent the luminous tail of this population.  Infrared
number counts and the cosmic infrared background provide strong
evidence that the ULIRG population becomes progressively more
important at high redshift \citep{Dole01, Elbaz02, LeFloch04}.  In addition,
submillimeter galaxies (SMGs) have been revealed as likely ULIRG
analogs at high redshift \citep{Smail97b, Barger98, Hughes98,
Ivison98, Eales00, Scott02}.  However, although the energy generation
mechanisms of these extremely luminous objects have been a topic of
great interest and study, the relative fraction of high redshift
SMGs powered by buried AGN is still a matter of debate, as are the
relative contributions of AGN and starbursts to the energy output of
ULIRGs at low redshift.

Mid-infrared diagnostics of the energetics of galaxies have been
developed using ground-based \citep{Roche91} and
space-based \citep{Genzel98,Lutz00,Laurent00} observations.
In general, the mid-infrared spectra of starbursts are characterized
by low-excitation fine structure lines, polycyclic aromatic
hydrocarbon (PAH) features, and a weak 3--6 $\micron$ rest-frame
continuum.  In contrast, the mid-infrared spectra of AGN exhibit
high-excitation emission lines, very weak or absent PAHs, and a strong 3--6 $\micron$
rest-frame continuum.  Until recently, the sensitivity of available
spectrometers has limited the application of these diagnostics to
the brightest galaxies at low redshift.  The highly sensitive instruments aboard the
\spitzer \ \citep{Werner04}, especially the \textit{Infrared
Spectrograph} \citep[\irs;][]{Houck04}, are powerful tools for
performing detailed studies of ULIRGs at higher redshifts than were
previously possible with \textit{IRAS} or \textit{ISO}.

In order to identify high-redshift ULIRGs for detailed study, the \irs
\ and \mips \ instrument teams have completed a mid-infrared imaging
survey of the 9 deg$^2$ \bootes \ region of the NOAO Deep Wide-Field
Survey (NDWFS; Jannuzi \etal, in preparation).  The NDWFS is a $B_WRIK$ imaging
survey reaching 3$\sigma$ point-source depths of approximately 27.7,
26.7, 26.0, and 19.6 Vega magnitudes, respectively.  In particular, we used the NDWFS Data Release 3 catalog.  The mid-infrared
imaging was carried out with the \textit{Multiband Imaging Photometer
for Spitzer} \citep[\mips;][]{Rieke04} and reaches 5$\sigma$ limits of
0.28, 35, and 100 mJy at 24, 70, and 160 $\micron$, respectively.
Most (8.5 deg$^2$) of this field has also been surveyed using the 
\textit{Infrared Array Camera} \citep[\irac;][]{Fazio04}, also aboard \textit{Spitzer}, to 5$\sigma$
sensitivities of 6.4, 8.8, 51, and 50 $\mu$Jy at wavelengths of 3.6,
4.5, 5.8, and 8.0 $\micron$, respectively \citep{Eisenhardt04}.  The
ACIS-I instrument \citep{Garmire03} aboard the \textit{Chandra X-ray
Observatory} was also used to map the entire \bootes \ field down to a
limiting sensitivity of $\sim 4 \times 10^{-15}$ erg cm$^{-2}$
s$^{-1}$ in the energy range 0.5--7 keV \citep{Murray05}.  The \bootes
\ field also overlaps with the 20 cm FIRST Survey, which has a
limiting sensitivity of approximately 1 mJy \citep{White97}.

The NDWFS \bootes \ field was chosen for the MIPS imaging survey
because of its low infrared background and high-quality, deep $B_WRI$
imaging.  These properties allow the selection of mid-infrared sources
which are bright enough for \irs \ spectroscopy but which have very faint
optical counterparts, and are therefore likely at high redshift.
\citet{Houck05} describe the results of \irs \
observations of 30 extreme ($f_{\nu}(24\micron) > 0.75$ mJy; $R > 24$
mag) sources identified in \bootes.  The vast majority of the 17
objects for which the mid-infrared spectra yielded redshifts are
probably obscured AGN at $z \approx 2$.  We present here a detailed
analysis of two objects which were also selected from the \bootes \
field, but which have unique infrared and optical properties.  In
particular, we present \irs \ and optical spectroscopy, and constrain
the spectral energy distributions of these objects using data from the
previously described multiwavelength \bootes \ surveys, as well as new
submillimeter data.

\firstobjectlong \ (hereafter \firstobject) was selected from a
catalog of objects that were detected in all three MIPS bands (thus
the prefix ``MIPS'').  Following the naming scheme of \citet{Borys05},
we use the NDWFS $I$-band coordinates\footnote{For completeness, the
24 $\micron$ coordinates are $14^{\rm h}28^{\rm m}24\fs07$,
+35$\degr$26$\arcmin$19\farcs14 (J2000).} to form the name of this object.
\firstobject \ was chosen for follow-up because it is bright at 160
$\micron$ ($f_{\nu}(160\micron) = 430 \pm 90$ mJy), has red
optical--near-infrared colors ($R-K > 5$), and is unresolved in the
optical.  The large 160 $\micron$ flux indicates that \firstobject \
contains large amounts of cold dust.  The red optical--near-infrared
colors and compact morphology suggest that it resides at $z > 1$.  The
combination of the 160 $\micron$ flux and the high redshift implies
that \firstobject \ probably has a large far-infrared luminosity
($\lfir \simeq 10^{13} {\rm L}_{\odot}$).  \citet{Borys05} presented
the spectral energy distribution (SED) of this object from the
optical through the radio.  They argued that while the emission
shortward of 1 $\micron$ (observed frame) is likely dominated by a
$z \approx 1$ lens, the near-infrared emission is dominated by the 24 $\micron$ source.  Based on a cool ($\sim$43 K) dust temperature, adherence
to the far-infrared--radio correlation, and the presence of a
prominent 1.6 $\micron$ stellar bump, they argued that \firstobject \
is a dusty starburst.  As discussed in \S \ref{sec:firstPAH}, the \irs
\ spectrum we present provides a rough redshift of $z \approx 1.3$.
\citet{Borys05} used this estimate to identify H$\alpha$ in the
near-infrared spectrum and obtained $z = 1.325 \pm 0.002$.  In
combination with their SED, this redshift confirms a high luminosity
of $\lfir = (3.2 \pm 0.7) \times 10^{13} \Lsun$.  They also noted that
the instrinsic luminosity may be up to a factor of 10 lower if
lensing is important.  Here we present additional diagnostics to
determine the energetics of \firstobject.

\secondobjectlong \ (hereafter \secondobject), was selected from a
catalog of objects detected at 24 $\micron$ (thus the prefix
``SST24'').  Following the convention of \citet{Houck05}, we use the
24 $\micron$ position to derive the name of this object.
\secondobject \ has the brightest 24 $\micron$ flux ($f_{\nu}(24
\micron) = 10.55$ mJy) among mid-infrared sources with faint optical
counterparts ($R > 20$).  The faint $R$-band magnitude ($R = 22.78$)
implies that \secondobject \ is at high redshift, and the 24 $\micron$ flux indicates a large luminosity.

The plan of this paper is as follows: \S \ref{sec:Observations}
describes the mid-infrared \irs \ observations of both sources, as well as optical spectroscopy and submillimeter
imaging of \secondobject.  The results for each object are presented
in \S \ref{sec:firstobjectlong} and \S \ref{sec:secondobjectlong}.
Finally, our conclusions are discussed in \S \ref{sec:Conclusions}.
Throughout, we use the following cosmological parameters: $\Omega_M =
0.3$, $\Omega_{\Lambda} = 0.7$, and $H_{0} = 70 $ km s$^{-1}$
Mpc$^{-1}$.

\section{Observations and Data Reduction}
\label{sec:Observations}

\subsection{\textit{Infrared Spectrograph (IRS)}}
\label{sec:irs}

Both objects were observed with the \irs \ Short Low (SL) module in
first order for a total of 480 seconds and with the Long Low (LL)
module in first and second orders for a total of 1920 seconds each.
Targets were placed on the slits by performing moderate-accuracy
peak-ups on and offsetting from nearby 2MASS stars.  Observed two
dimensional spectra were processed with version S11.0 of the Spitzer
Science Center pipeline.  Further processing was carried out using the
Basic Calibrated Data (BCD) products from the pipeline.  Since the SL
data were taken in first order only, a background image for each
position (nod) was constructed from data taken in the other position.
In contrast, the background images for each LL position were
constructed from data taken when the object was in the other order.
Extraction of one dimensional spectra corresponding to each position
was accomplished with the SMART analysis package \citep{Higdon04}.
The data from each position were then averaged to produce the final
spectra, which provide low resolution coverage from $\sim$8--38
$\micron$.

\subsection{\textit{DEIMOS}}
\label{sec:deimos}

Optical spectra of both \firstobject \ and \secondobject \ were
obtained with the Deep Imaging Multi-Object Spectrograph
\citep[DEIMOS;][]{Faber03} on the W. M. Keck II 10-meter telescope on
the night of UT 2005 May 06. The observations, three 30 minute
exposures, were obtained in ~0.5" seeing through a 1.0" wide and 10"
long slitlet oriented at PA=130. A 600 line mm$^{-1}$ (7500\AA \
blaze) grating setting with the GG400 blocking filter was used. The
central wavelength was set to 7500\AA. This corresponds to a 0.65\AA \
per pixel mean spectral dispersion.

The DEEP2 data reduction pipeline\footnote{{\tt
http://alamoana.keck.hawaii.edu/inst/deimos/pipeline.html}} was used
to perform cosmic ray removal, flat-fielding, co-addition,
sky-subtraction, and wavelength calibration. The one-dimensional
spectra were extracted using standard IRAF routines. Relative
spectrophotometric calibration was performed using observations of
Wolf 1346 \citep{Massey88,Massey90,Oke90}.

The optical spectrum of \firstobject \ is presented and discussed by
\citet{Borys05}.

\subsection{SHARC-II}
\label{sec:sharcII}

Submillimeter images of \secondobject \ were taken on UT 2004 April
10--12 with the 350 $\micron$ filter on the SHARC-II camera at the
Caltech Submillimeter Observatory.  Pointing and flux calibration were
performed on nearby Arp 220, and data were reduced using the CRUSH
software package \citep{Kovacs05}.  Weather conditions were not
optimal, with an average opacity of $\tau_{\rm 225\,GHz}\sim0.07 $.
The total exposure time was roughly 1 hour.  The object was not
detected, though we derive a 3$\sigma$ upper limit of 150 mJy.

\begin{deluxetable*}{ccccc}
\tablecaption{Multiwavelength Photometry}
\tabletypesize{\scriptsize}
\tablewidth{0pt}
\startdata \hline\hline
                                 & \multicolumn{2}{c}{\firstobjectlong \ (\firstobject)}   & \multicolumn{2}{c}{\secondobjectlong}  \ (\firstobject)\\
Wavelength                       & \multicolumn{1}{c}{Flux}                      & \multicolumn{1}{c}{Instrument\tablenotemark{a}}             & \multicolumn{1}{c}{Flux\tablenotemark{b}}   & \multicolumn{1}{c}{Instrument\tablenotemark{a}} \\ 
\hline
445 nm ($B_W$)                    & 0.25  $\pm$ 0.03  $\mu$Jy & MOSAIC-I 	       & 0.98 $\pm$ 0.03 $\mu$Jy       &  MOSAIC-I               \\
658 nm ($R$)                      & 2.02  $\pm$ 0.08  $\mu$Jy & MOSAIC-I 	       & 2.29 $\pm$ 0.10 $\mu$Jy       &  MOSAIC-I               \\
806 nm ($I$)                      & 6.21  $\pm$ 0.11  $\mu$Jy & MOSAIC-I 	       & 5.21 $\pm$ 0.13 $\mu$Jy       &  MOSAIC-I               \\
1.22 $\micron$ ($J$)              & 31.9 $\pm$ 5.3 $\mu$Jy    & WIRC                   & \nodata                       & \nodata                 \\
1.63 $\micron$ ($H$)              & 38.4 $\pm$ 5.2 $\mu$Jy    & WIRC                   & \nodata                       & \nodata                 \\
2.19 $\micron$ ($K$)              & 72.4  $\pm$ 9.4   $\mu$Jy & WIRC     	       & 134.8 $\pm$ 10.6 $\mu$Jy      &  ONIS                   \\
3.6 $\micron$                     & 250.9 $\pm$ 7.2   $\mu$Jy & IRAC     	       & 296.3 $\pm$ 8.9 $\mu$Jy       &  IRAC                   \\
4.5 $\micron$                     & 280.6 $\pm$ 8.7   $\mu$Jy & IRAC    	       & 621.3 $\pm$ 18.6 $\mu$Jy      &  IRAC                   \\
5.8 $\micron$                     & 198.3 $\pm$ 16.6  $\mu$Jy & IRAC     	       & 1600.9 $\pm$ 48.0 $\mu$Jy     &  IRAC                   \\
8.0 $\micron$                     & 211.1 $\pm$ 14.2  $\mu$Jy & IRAC     	       & 3938.9 $\pm$ 118.2 $\mu$Jy    &  IRAC                   \\
24 $\micron$                      & 0.72  $\pm$ 0.07  mJy & MIPS     	               & 10.55 $\pm$ 0.13 mJy          &  MIPS                   \\
70 $\micron$                      & 34    $\pm$ 6     mJy     & MIPS     	       & \nodata                       &  \nodata                \\
160 $\micron$                     & 430   $\pm$ 90    mJy     & MIPS     	       & $<$45 mJy (3$\sigma$)         &  MIPS                   \\
350 $\micron$                     & 226   $\pm$ 45    mJy     & SHARC-II 	       & $<$150 mJy (3$\sigma$)        &  SHARC-II               \\
850 $\micron$                     & 21.9  $\pm$ 1.3   mJy     & SCUBA    	       & \nodata                       &  \nodata                \\
20 cm                             & 0.937 $\pm$ 0.039 mJy     & VLA\tablenotemark{c}   & $<$0.98 mJy (5$\sigma$)       &  VLA\tablenotemark{d}   \\

\enddata
\tablecomments{The optical, NIR, and IRAC fluxes are measured in 5$\arcsec$ diameter apertures.  Total fluxes are reported for the longer wavelength bands.  The fluxes for \firstobjectlong \ were taken from \citet{Borys05}.}
\tablenotetext{a}{MOSAIC-I is an imager at the KPNO 4m, WIRC is the NIR camera at Palomar, SHARC-II is at the CSO, SCUBA is at the JCMT, and ONIS was previously available at the KPNO 2.1m.}
\tablenotetext{b}{Missing values indicate that no data were taken at the given wavelength.  In particular, although most of the \bootes \ field was imaged at 70 $\micron$, \secondobject \ lies near the edge of the field, outside of the 70 $\micron$ coverage.  Limits are provided where data were taken, but no source was detected.}
\tablenotetext{c}{\citet{Higdon05}}
\tablenotetext{d}{FIRST Survey; \citet{White97}}
\label{table:MultiwavelengthPhotometry}
\end{deluxetable*}

\section{\firstobjectlong \ (\firstobject)}    
\label{sec:firstobjectlong}

\subsection{PAH Emission Bands}
\label{sec:firstPAH}

\begin{figure*}
\epsscale{0.9}
\plotone{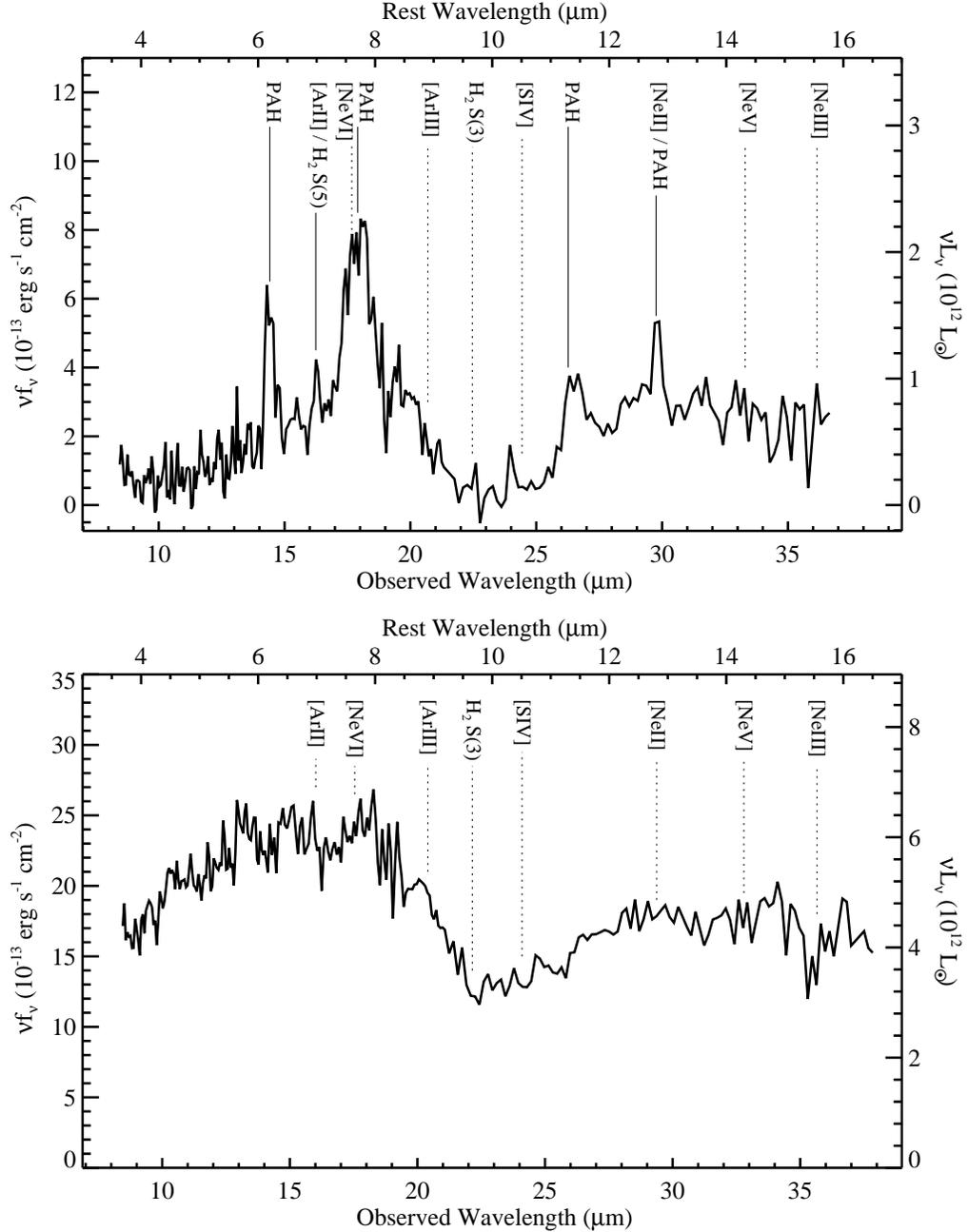}
\caption{\irs \  spectra of \firstobjectlong \ (\firstobject; top) and \secondobjectlong \ (\secondobject; bottom).  Features labeled with solid lines are detected, while those labeled with dotted lines are undetected.  Note that the narrow 12.8 $\micron$ \ion{Ne}{2} emission lies on top of the broader 12.7 $\micron$ PAH feature.}
\label{fig:spectra}
\end{figure*}

As seen in the top panel of Figure \ref{fig:spectra}, the mid-infrared
spectrum of \firstobject \ is rich in emission features, including the
PAH bands at 6.2, 7.7, 11.3, and 12.7 $\micron$. These features
indicate a redshift of $z = 1.34 \pm 0.02$.  Based on a preliminary
reduction of the \irs \ data, \citet{Borys05} identified the narrow
H$\alpha$ and adjacent [\ion{N}{2}] lines in a near-infrared spectrum
to determine a more accurate redshift of $z=1.325 \pm 0.002$.  All
plots and calculations herein adopt this more accurate redshift.
Based on the optical spectrum, \citet{Borys05} also considered whether or
not the mass of a foreground $z \approx 1$ object in the field is
significantly magnifying the background $z \approx 1.3$ source.
Luminosities have not been corrected, since the magnification is
unknown.
 
The PAH features in the \irs \ spectrum of \firstobject \ not only
provide its redshift, but are also clues to the source of its energy
generation.  The strength of PAH features roughly correlates with the
starburst contribution to the mid-infrared flux \citep[][ Armus \etal,
in preparation]{Genzel98,Lutz98,Rigopoulou99,Laurent00}.  In the
spectrum of \firstobject, the 7.7 and 11.3 $\micron$ PAH features are
adjacent to the broad 9.7 $\micron$ silicate absorption band, making
the continuum at these wavelengths uncertain, while the 12.7 $\micron$
PAH feature is blended with the [\ion{Ne}{2}] emission line at 12.8
$\micron$.  We therefore use the 6.2 $\micron$ PAH rest-frame
equivalent width (${\rm EW}_{\rm rest}({\rm PAH}6.2\micron)$) as our
primary diagnostic of the starburst contribution to the energy
generation.

For \firstobject, ${\rm EW}_{\rm rest}({\rm PAH}6.2\micron) = 0.37 \pm
0.04 \ \micron$, 25--50\% lower than the typical value measured for
nearby starbursts (0.5 -- 0.7 $\micron$; Brandl \etal, in
preparation), but much larger than that measured for nearby AGN
\citep[0.005 -- 0.02 $\micron$;][]{Weedman05}.  The equivalent width
diagnostic therefore indicates that the mid-infrared spectrum of
\firstobject \ is starburst-dominated.

\begin{figure*}
\epsscale{0.75}
\plotone{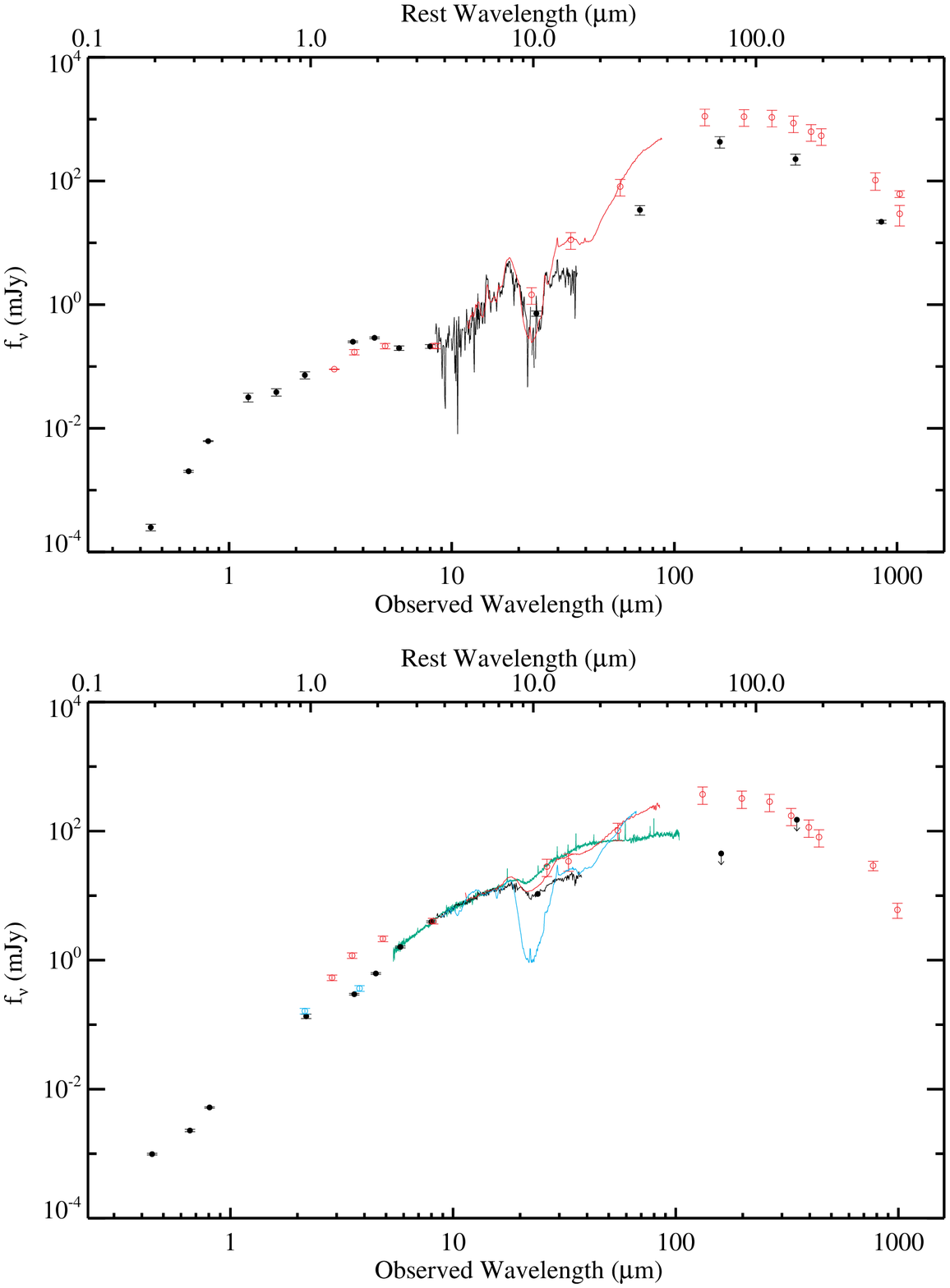}
\caption[SEDs of \firstobject \ and \secondobject]{\textit{Top:}  The observed spectral energy distribution (SED) of \firstobjectlong \ (\firstobject) is plotted with black lines and black filled circles.  The SED of Arp220, normalized at 7.7 $\micron$, is overplotted with red lines and red hollow circles.  \textit{Bottom:}  The SED of \secondobjectlong \ (\secondobject) is plotted with black lines and black filled circles, and the SEDs of F00183-7111 \citep{Spoon04}, Mrk231 \citep{Weedman05}, and NGC1068 \citep{Sturm00}, normalized at 13--14 $\micron$, are overplotted with blue, red, and green lines and hollow symbols.}
\label{fig:seds} 
\end{figure*}

Compared to local ULIRGs (Armus \etal, in preparation), \firstobject \
has a large 6.2 $\micron$ PAH luminosity of $\lsix = (3.0 \pm 0.3)
\times 10^{10} {\rm L}_{\odot}$.  However, the top panel of Figure
\ref{fig:seds} shows that this object has strong $160~\micron$
emission from large amounts of $\sim$43K dust.  We therefore
quantified the contribution of the PAH feature to the bolometric
luminosity of the source.  For \firstobject, $\lsix / \lfir = (9.4 \pm
2.3) \times 10^{-4}$, where the error bar includes contributions from
the uncertainties in both $\lsix$ and $\lfir$.  The latter was
determined by \citet{Borys05}, and includes uncertainties associated
with fitting a modified blackbody to the SED.  It does not include an
estimate of the range of alternate models that could be used for this
fit.  Since source structure depends on wavelength, differential
magnification as a function of wavelength represents an added
complication in measuring $\lsix / \lfir$.

The value of $\lsix / \lfir$ measured for \firstobject \ is a factor
of $\sim$2--3 smaller than that seen in other prototypical starbursts
and starburst-dominated ULIRGs, including NGC7714, M82, and UGC5101,
which have $\lsix / \lfir = 26 \times 10^{-4}$, $20 \times 10^{-4}$,
and $16 \times 10^{-4}$, respectively.  It is within a factor of 1.5
smaller than the values measured for the $z \approx 2$
starburst-dominated ULIRGs recently discovered by \citet{Yan05}, which
have $\lsix / \lfir = 11 \times 10^{-4}$ and $13 \times 10^{-4}$.
Given the substantial uncertainties discussed above, the normalized
PAH luminosity of \firstobject \ is consistent with all of these
comparison objects.

Recently, \citet{Lutz05} presented the \irs \ spectra of two $\lfir
\ga 10^{13} {\rm L}_{\odot}$ SMGs at $z \simeq 2.8$. They found that
these SMGs appear to be scaled-up versions of local ULIRGs, based on a
comparison of their PAH to far-infrared ratios\footnote{defined as the
ratio of the peak flux density of the continuum-subtracted 7.7
$\micron$ PAH feature to the rest-frame 222 $\micron$ (observed-frame
850 $\micron$ at $z\sim2.8$) continuum flux density}.  Using the
best-fitting modified blackbody derived for \firstobject \ by
\citet{Borys05}, we find its PAH to far-infrared ratio is
$\log(f_{\nu}({\rm PAH} 7.7\micron) / f_{\nu}(222\micron)) = -1.28$.
This is very similar to the values reported for the $z \simeq 2.8$
SMGs and local ULIRGs.

\subsection{Emission Lines}
\label{sec:firstlines}

In addition to PAH features, the mid-infrared spectra of galaxies may
contain additional emission features, such as [\ion{Ar}{2}] at 6.9
$\micron$, [\ion{Ar}{3}] at 8.9 $\micron$, [\ion{S}{4}] at 10.5
$\micron$, [\ion{Ne}{2}] at 12.8 $\micron$, [\ion{Ne}{3}] at 15.6
$\micron$, [\ion{Ne}{6}] at 7.6 $\micron$, and [\ion{Ne}{5}] at 14.3
$\micron$ (all rest wavelengths).  The expected positions of these features are marked in
Figure \ref{fig:spectra}.  Due in part to the low signal-to-noise and
resolution of our spectra, many of these lines are not detected.
However, it is possible that a blend of [\ion{Ar}{2}] and H$_2$ S(5)
is detected near 7 $\micron$, and [\ion{Ne}{2}] at 12.8 $\micron$ is
clearly seen on top of the broader PAH feature, with a flux of $(2.8
\pm 0.3) \times 10^{-18}$ W m$^{-2}$.  We measure a 3$\sigma$ lower
limit of [\ion{Ne}{2}]/[\ion{Ne}{3}] $>$ 1.24, a high value which is
consistent with those predicted and measured for starbursts
\citep{Spinoglio92,Thornley00}.  Because OB stars cannot ionize
\ion{Ne}{4}, the detection of the 14.3 $\micron$ [\ion{Ne}{5}] emission
line would imply the presence of an AGN.  Although we do not detect this
line, our 3$\sigma$ upper limit ($2.14 \times 10^{-18}$ W m$^{-2}$)
implies [\ion{Ne}{5}] 14.3 $\micron$ / [\ion{Ne}{2}] 12.8 $\micron <
0.76$.  It is not uncommon for AGN to exhibit similar ratios
\citep{Weedman05}, so this limit cannot rule out the presence of an
AGN in \firstobject.

\subsection{Absorption}
\label{sec:firstabsorption}

The mid-infrared spectrum of \firstobject \ displays a broad silicate
absorption trough near 9.7 $\micron$.  It is difficult to accurately
estimate a silicate optical depth for a PAH-dominated spectrum because
of the uncertainty in determining the continuum level.  Using a simple
procedure, we assume a power-law continuum between the measured flux
densities at 5.5 and 14.5 $\micron$ (rest-frame).  The 5.5 $\micron$
continuum anchor point lies sufficiently shortward of the 6.2
$\micron$ PAH feature to remain free of PAH emission.  In addition,
\firstobject \ shows no evidence of water ice or hydrocarbon absorption,
which are sometimes seen in ULIRG spectra at these wavelengths.  Our
long wavelength continuum anchor is well beyond both the 12.7
$\micron$ PAH feature and the (unobserved) 12.8 $\micron$
[\ion{Ne}{2}] emission line.  At 14.5 $\micron$, ULIRG spectra are
dominated by thermal emission from warm, small grains (see \S
\ref{sec:ContinuumSlope}). This procedure yields an optical depth of
$\tau(9.7 \micron) \ga 1.5$, presented as a lower limit because of the
unknown extinction at the chosen continuum points.  Adopting the
extinction law of \citet{Draine03}, our limit on $\tau(9.7 \micron)$
implies $A(V) \ga 27.8$ mag.  This large visual extinction may explain
why the optical spectrum indicates a different redshift than the
mid-infrared spectrum \citep{Borys05}.

\subsection{Continuum at $\lambda > 8 \micron$}
\label{sec:ContinuumSlope}

\begin{figure*}
\centerline{
\psfig{figure=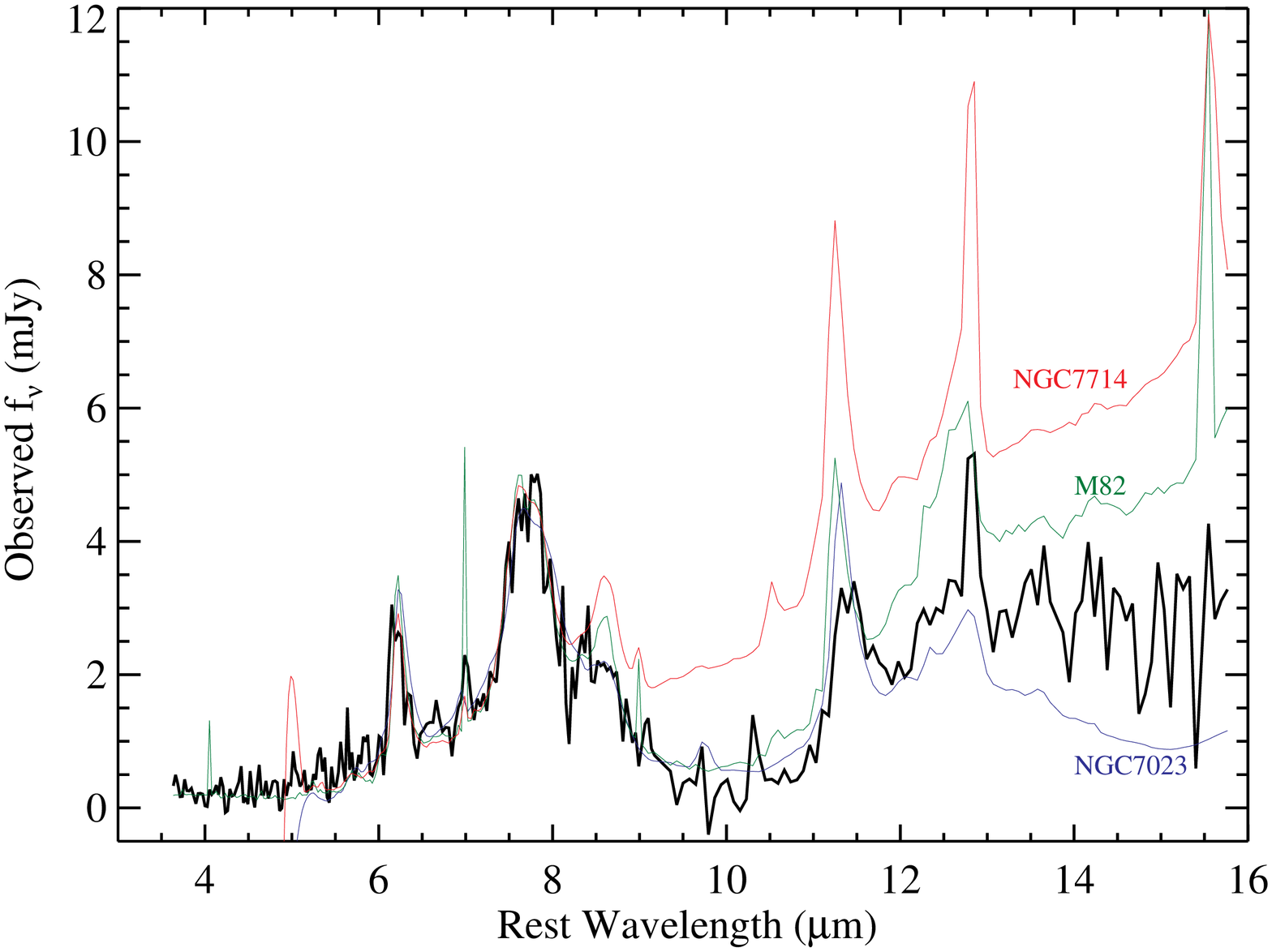,width=6truein,angle=270}}
\caption[Mid-infrared spectra of \firstobject \ and comparison sources]{The mid-infrared \irs \ spectrum of \firstobjectlong \ (\firstobject; thick line) compared with NGC7023, M82, and NGC7714 (thin lines). The latter three have been normalized at rest-frame 7.7 $\micron$ to match the observed flux density of \firstobjectlong.}
\label{fig:absorption} 
\end{figure*}

The mid-infrared spectra of galaxies can be modeled as combinations of
the spectra of \ion{H}{2} regions; the photo-dissociation regions
(PDRs) around \ion{H}{2} regions; and AGN-heated dust
\citep{Laurent00}.  As we showed in \S \ref{sec:firstPAH},
\firstobject \ is powered primarily by a starburst.  Thus, we would
expect its mid-infrared spectrum to be dominated by PAH-rich PDRs and
\ion{H}{2} regions, which typically exhibit steeply rising continua at
12--16 $\micron$ due to very small grains with radii less than 10 nm
\citep[VSGs;][]{Desert90}.  In Figure \ref{fig:absorption}, we compare
the mid-infrared spectrum of \firstobject \ with those of the
unobscured Galactic reflection nebula NGC7023 \citep[pure
PDR;][]{Werner04a} and the prototypical starbursts NGC7714
\citep{Brandl04} and M82 \citep{Sturm00}.  All three spectra are
similar at rest-wavelengths shortward of 8 $\micron$.  Like most
starbursts, NGC7714 and M82 have strong VSG continua which set in longward of 8 $\micron$.  However, the mid-infrared spectrum of
\firstobject \ has only a weak VSG continuum.

\subsection{Multiwavelength Photometry}
\label{sec:firstphotometry}
A large hard X-ray flux can indicate the presence of an AGN.
\firstobject \ was not detected in a 5 ks Chandra survey of the
\bootes \ field \citep{Murray05}.  This non-detection corresponds to
an upper limit of $L_{2-10 {\rm keV}} / \lfir < 0.001$ for neutral
Hydrogen column densities of $N_H < 10^{23}$ cm$^{-2}$, assuming a
Galactic column density of $N_H = 10^{20}$ cm$^{-2}$ and a power law
spectrum with photon index $\Gamma = 1.7$.  This limit is consistent
with the values of $L_{2-10 {\rm keV}} / \lfir$ measured for
starburst-dominated galaxies \citep{Ptak03}.  However, we cannot rule
out the presence of a Seyfert-like nucleus behind neutral Hydrogen
column densities of $\ga 10^{24}$ cm$^{-2}$.

The shape of the SED also provides information on the power source.
\citet{Borys05} presented the optical to radio SED of
\firstobject, and argued that it is a starburst based upon its cool
dust temperature, adherence to the far-infrared--radio correlation,
and the prominence of the 1.6 $\micron$ stellar bump. Using a large
spectroscopically-classified sample, \citet{Stern05} have developed
\irac \ color criteria for identifying AGN.  Interestingly, by these standards, \firstobject \ would
be classified as an AGN.  However, those authors point out that their
criteria may result in contamination from $z \approx 1.4$ ULIRGs.

\pagebreak

\section{\secondobjectlong \ (\secondobject)}
\label{sec:secondobjectlong}

\begin{figure*}
\centerline{
\psfig{figure=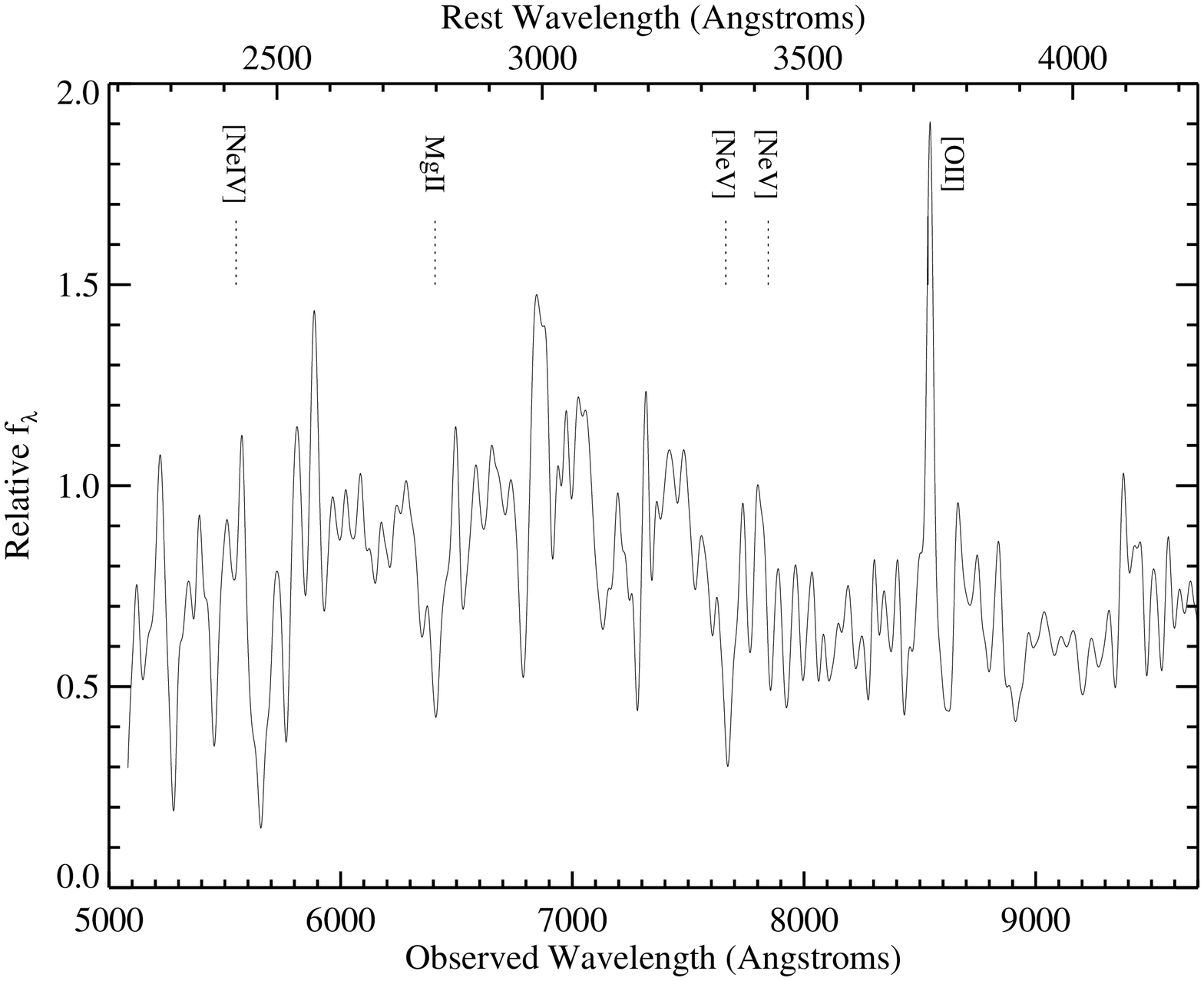,width=6truein,angle=270}}
\caption[Optical spectrum of \secondobjectlong]{Optical spectrum of \secondobjectlong \ (\secondobject), taken with the DEIMOS spectrograph at the Keck II telescope and smoothed with a Gaussian filter.  The [\ion{O}{2}]3727\AA \ emission line is clearly detected at 8545\AA, providing a redshift of $z=1.293 \pm 0.001$.  The wavelengths of additional features which are sometimes found in galaxy spectra are indicated by dotted lines.  None are detected in \secondobject.}
\label{fig:deimos} 
\end{figure*}

The \irs \ spectrum of \secondobject, shown in the bottom panel of
Figure \ref{fig:spectra}, differs dramatically from that of
\firstobject.  The only obvious feature is the broad 9.7 $\micron$
silicate absorption band.  Assuming a power-law continuum between 7.8
and 14.5 $\micron$ (rest-frame), we obtain an optical depth of
$\tau(9.7 \micron) \ga 0.56$.  This corresponds to $A(V) \ga 10.3$ mag
\citep{Draine03}.  As with \firstobject, these are lower limits due to
the unknown extinction at the continuum points.  The placement of the
absorption feature also indicates a redshift of $z \approx 1.3$.  We
obtained a more accurate redshift of $z = 1.293 \pm 0.001$ from the
[\ion{O}{2}]3727\AA \ emission line in a Keck-DEIMOS optical spectrum
(see Figure \ref{fig:deimos}).  Unfortunately, no optical
spectroscopic diagnostics of the energetics of \secondobject \ could
be identified.

\secondobject \ has an extremely large mid-infrared luminosity.  At 5
$\micron$ (rest-frame), it has $\nu {\rm L}_{\nu} = 5.5 \times
10^{12}~\Lsun$.  Out of the 17 objects with redshifts from the
\citet{Houck05} sample, only one (SST24J143001.91+334538.4 at $z =
2.46$) has a comparable 5 $\micron$ luminosity.  The remainder were
either not sampled at 5 $\micron$ (4 sources) or are at least a factor
of $\sim$2 less luminous.  The 5 $\micron$ luminosity of \secondobject
\ is also over an order of magnitude brighter than any of the six $z >
1$ ULIRGs observed with the \irs \ by \citet{Yan05} and at least a
factor of three brighter than even the most luminous local ULIRGs.

The powerful soft X-ray and ultraviolet radiation associated with an
AGN may destroy PAH carriers \citep[e.g.][]{Aitken85,Voit92}.  The
lack of PAH features in \secondobject \ (${\rm EW}_{\rm rest}({\rm
PAH}6.2\micron) < 0.014$ $\micron$; 3$\sigma$) suggests that its
luminosity is generated primarily by an AGN, since the PAH equivalent
width is a factor of $\sim$30--40 lower than found in pure starbursts
(Brandl \etal, in preparation).

The SED of \secondobject \ is shown in black in the bottom panel of
Figure \ref{fig:seds}.  The 8--1000 $\micron$ luminosity is poorly
constrained by the available data.  However, given its extremely high
mid-infrared luminosity, it is likely that \secondobject \ has $\lfir
\ga 10^{13} {\rm L}_{\odot}$.  In addition, the optical and
near-infrared data do not display a
stellar bump, suggesting that this feature is swamped by a hot dust continuum produced by an AGN.  For reference, we
have overplotted the normalized spectra of F00183-7111
\citep{Spoon04}, Mrk231 \citep[][ Armus \etal, in
preparation]{Weedman05}, and the nucleus of NGC1068 \citep{Sturm00}.
All three objects are probably AGN-dominated with energetically
significant circumnuclear star formation, but the AGN-heated dust in
F00183-7111 and Mrk231 is deeply obscured, and the NGC1068 spectrum is
dominated by nuclear, rather than circumnuclear, emission.
\secondobject \ shows a much smaller silicate absorption than
F00183-7111, and its 160 and 350 $\micron$ limits indicate that it has
smaller relative amounts of cool dust than either F00183-7111 or
Mrk231.  Its SED is most similar to that of the nucleus of NGC1068,
where we have a direct line-of-sight to the hot dust heated by the
AGN.

\secondobject \ was not detected in the 5 ks Chandra survey, but we
are unable to put useful limits on $L_{2-10 {\rm keV}} / \lfir$
because $\lfir$ is poorly constrained.  It was also undetected in
the 20 cm FIRST Survey \citep{White97}, providing a 5$\sigma$ upper
limit of 0.98 mJy.

\section{Conclusions}
\label{sec:Conclusions}

We have presented the \irs \ spectra of two $z \approx 1.3$ ULIRGs
selected from the \bootes \ region of the NDWFS.  

\firstobject \ has a high 6.2 $\micron$ PAH equivalent width, high
[\ion{Ne}{2}]/[\ion{Ne}{3}] ratio, and low ${\rm L}_{2-10{\rm keV}}/\lfir$.
These properties, in combination with the prominent 1.6 $\micron$
stellar bump, cold ($\sim$43 K) dust temperature, and concordance with
the radio-FIR correlation noted by \citet{Borys05}, indicate that
\firstobject \ is dominated by a starburst.  With $\lfir = 3.2 \times
10^{13} \Lsun$ (modulo its unknown lensing amplification),
\firstobject \ is extremely luminous compared to low-redshift
starbursts.  However, PAH-dominated spectra of comparably luminous
objects are beginning to be discovered at $z \ga 2$ \citep{Lutz05,
Yan05}.  \firstobject \ may be analogous to these more distant
luminous starbursts, but is at a redshift where it is more easily
studied in detail.  Its use as a template for higher redshift
populations rests on our ability to understand its lensing properties
\citep{Borys05}, which will require high-resolution imaging to
separate out the contribution from the foreground object.

At a similar redshift, \secondobject \ is among the most luminous
mid-infrared sources known (as measured at 5 $\micron$, rest-frame).
It appears to be powered mainly by an AGN: it has a very small 6.2
$\micron$ PAH equivalent width; exhibits a mid-infrared SED similar to
those of AGN-dominated ULIRGs; and lacks large amounts of cool dust,
as evidenced by its moderate silicate optical depth and non-detections
at 160 and 350 $\micron$.

The first sizable high-redshift samples of dusty sources observed with
the \irs \ have focused on objects at $z \ga 2$ \citep{Houck05,
Yan05}.  The submillimeter galaxies also lie at this redshift.
Relatively less is known about the population of dusty sources at $1 <
z < 2$. This redshift range has been traditionally difficult to access
through optical spectroscopy, and has come to be known as the
``redshift desert''.  Indeed, the optical spectra of both objects and
the near-infrared spectrum of \firstobject \ show very few emission
lines.  A redshift identification for either source would have been
dubious without its mid-infrared \irs \ spectrum.  Nevertheless, a
large fraction of the stars in local L$_{\ast}$ galaxies were likely
formed in this redshift range \citep[e.g.][]{Dickinson03,Drory05}.
Objects such as \firstobject \ and \secondobject \ provide examples of
how the \irs \ will allow the identification and detailed study of
objects in this critical redshift range.

\acknowledgments

We are thankful for enlightening discussions with Lin Yan and
thoughtful comments by the anonymous referee.

This work made use of images and data products provided by the NOAO
Deep Wide-Field Survey \citep[][ Dey \etal, in preparation; Jannuzi \etal, in preparation]{Jannuzi99}, which is
supported by the National Optical Astronomy Observatory (NOAO). NOAO
is operated by AURA, Inc., under a cooperative agreement with the
National Science Foundation.

The analysis pipeline used to reduce the DEIMOS data was developed at
UC Berkeley with support from NSF grant AST-0071048.

The authors wish to recognize and acknowledge the very significant
cultural role and reverence that the summit of Mauna Kea has always
had within the indigenous Hawaiian community.  We are most fortunate
to have the opportunity to conduct observations from this mountain.


{\it Facilities:} Spitzer(\irs, \mips, \irac), Keck:II (DEIMOS), Mayall (MOSAIC-I), KPNO:2.1m (ONIS), VLA, CSO (SHARC-II), JCMT (SCUBA)

\end{document}